\documentclass{article}
\usepackage{times}
\begin{document}
\title{Quantum Mechanics in Infinite Symplectic Volume}
\author{Jos\'e M. Isidro\\
Instituto de F\'{\i}sica Corpuscular (CSIC--UVEG)\\
Apartado de Correos 22085, Valencia 46071, Spain\\
{\tt jmisidro@ific.uv.es}}

\maketitle

\begin{abstract}
We quantise complex, infinite--dimensional projective space ${\bf CP}({\cal H})$.
We apply the result to quantise a complex, finite--dimensional, classical phase space 
${\cal C}$ whose symplectic volume is infinite, by holomorphically embedding it into 
${\bf CP}({\cal H})$. The embedding is univocally determined by requiring it to be an isometry 
between the Bergman metric on ${\cal C}$ and the Fubini--Study metric on ${\bf CP}({\cal H})$.
Then the Hilbert--space bundle over ${\cal C}$ is the pullback, by the 
embedding, of the Hilbert--space bundle over ${\bf CP}({\cal H})$. 


2001 Pacs codes: 03.65.Bz, 03.65.Ca, 03.65.-w. 


\end{abstract}

\tableofcontents

\hyphenation{di-men-sion}
\hyphenation{di-men-sio-nal}

\section{Introduction}\label{inntt}

\subsection{Notations}\label{nta}

Throughout this article, ${\cal C}$ will denote a complex $n$--dimensional, connected, 
classical phase space, endowed with a symplectic form $\omega$ 
and a complex structure ${\cal J}$. We will assume that $\omega$ and ${\cal J}$ 
are compatible, so holomorphic coordinate charts on ${\cal C}$ will also 
be Darboux charts. Upon quantisation, ${\cal H}$ will denote a complex, 
separable Hilbert space of quantum states. We will assume ${\cal C}$ to have an infinite 
symplectic volume,
\begin{equation}
\int_{\cal C}\omega^n = \infty,
\label{isxvcc}
\end{equation}
so ${\cal H}$ will be infinite--dimensional. Vector bundles over ${\cal C}$ with fibre 
${\cal H}$ will be called {\it quantum Hilbert--space bundles}, or ${\cal QH}$--bundles 
for short, and denoted ${\cal QH}({\cal C})$.

\subsection{Summary of results}\label{ssmmrr}

Classical phase space is the space of all solutions to the classical equations 
of motion (modulo gauge transformations that may eventually exist). As such it provides 
a natural starting point for quantisation. Although quantisation is a well--established procedure, 
recent developments in string and M--theory call for a revision of our understanding of 
{\it classical}\/ vs. {\it quantum} \cite{VAFA}. With this aim in mind we have analysed 
the quantisation of a complex, compact ${\cal C}$ in ref. \cite{PQM}. Compactness ensures that 
the symplectic volume of ${\cal C}$ is finite and, therefore, the Hilbert space of quantum states 
is finite--dimensional. Based on the example when ${\cal C}={\bf CP}^n$ we have analysed 
the different equivalence classes of (finite--dimensional) Hilbert--space bundles over ${\cal C}$. 

It remains to study the case when ${\cal C}$ has an infinite symplectic volume,
which we do in the present article extending the technique presented in ref. \cite{PQM}.
This corresponds, {\it e.g.}, to phase spaces such as ${\cal C}=T^*{\cal Q}$, the cotangent 
bundle to some configuration space ${\cal Q}$. The latter may well be compact, but its 
cotangent bundle is definitely noncompact. (Noncompactness of ${\cal C}$ is a necessary, 
though not sufficient, condition for its symplectic volume to be infinite). 
Of course, the quantisation of $T^*{\cal Q}$ has been known for long; there is a 
Hilbert space ${\cal H}$ where quantum states are square--integrable functions 
on ${\cal Q}$. We do not intend to alter this picture. Rather, our aim is 
to generalise it using the language fibre bundles over ${\cal C}$. 
In this language, the picture described above usually corresponds to the trivial bundle 
${\cal H}\times T^*{\cal Q}$. The notion of quantum--mechanical duality, as elaborated 
in ref. \cite{PQM}, suggests considering nonflat (and therefore nontrivial) Hilbert--space 
bundles over ${\cal C}$. This is what we do here.

Our approach may be summarised as follows. We first quantise a particular example of a complex, 
noncompact ${\cal C}$, the infinite--dimensional complex projective space ${\bf CP}({\cal H})$. 
Being infinite--dimensional, it cannot be the phase space of a finite number of degrees of freedom. 
However ${\bf CP}({\cal H})$ is easily quantised following ref. \cite{PQM}. For this
we consider a holomorphic atlas on ${\bf CP}({\cal H})$. Over each coordinate chart 
we erect a certain vector--space fibre; fibres are patched together across overlapping charts 
according to a set of transition functions. In this way a fibre bundle is constructed whose fibre 
is a Hilbert space of quantum states. The latter fall into two categories. 
One is the vacuum, the other its excitations. 
It turns out that all quantum states (except the vacuum) are holomorphic tangent vectors 
to ${\bf CP}({\cal H})$. This identifies the transition functions as jacobian matrices, 
and the corresponding subbundle as the holomorphic tangent bundle $T({\bf CP}({\cal H}))$. 
The vacuum $\vert 0\rangle_l$ appears as the fibrewise generator of a holomorphic line bundle 
$N_l({\bf CP}({\cal H}))$, one for each $l\in{\bf Z}$. The latter is the
Picard group of ${\bf CP}({\cal H})$. The complete quantum Hilbert--space bundle ${\cal
QH}_l({\bf CP}({\cal H}))$ is
\begin{equation}
{\cal QH}_l({\bf CP}({\cal H}))=T({\bf CP}({\cal H}))\oplus N_l({\bf CP}({\cal
H})),\quad l\in{\bf Z}.
\label{compelto}
\end{equation}

Next an $n$--dimensional, complex phase space ${\cal C}$ with infinite symplectic 
volume is quantised by holomorphically embedding it into ${\bf CP}({\cal 
H})$. The embedding $\iota\colon {\cal C}\rightarrow {\bf CP}({\cal H})$ 
is univocally determined by the condition that it be an isometry between 
the Bergman metric on ${\cal C}$ and the Fubini--Study metric on ${\bf CP}({\cal H})$. 
Then the bundle ${\cal QH}_l({\cal C})$ is the pullback by $\iota$ of the 
bundle ${\cal QH}_l({\bf CP}({\cal H}))$. 
In particular, ${\cal QH}_l({\cal C})$ is infinite--dimensional as required by the 
infinite symplectic volume of ${\cal C}$.

The differential geometry of ${\bf CP}({\cal H})$ was studied long ago in ref. \cite{CPINF},
and its deformation quantisation more recently in the nice paper \cite{MEX}. 
We have addressed duality in quantum mechanics from different perspectives 
in previous publications \cite{MEPREV}. For background material see, 
{\it e.g.}, refs. \cite{KN, LIBAZCA, LIBSCHLICHENMAIER, BRUZZO}. 
Finally we would also like to mention refs. \cite{MATONE, ANANDAN, FUJII}.

\section{The space of rays in Hilbert space}\label{innter}

Realise ${\cal H}$ as the space of infinite sequences of complex numbers $Z^1, Z^2, \ldots$ that 
are square--summable, $\sum_{j=1}^{\infty}\vert Z^j\vert ^2 < \infty$. 
The $Z^j$ provide a set of holomorphic coordinates on ${\cal H}$. 
The space of rays ${\bf CP}({\cal H})$ is
\begin{equation}
{\bf CP}({\cal H})=({\cal H}-\{0\})/({\bf R}^+\times U(1)).
\label{cph}
\end{equation}

The $Z^j$ provide a set of {\it projective}\/ coordinates on ${\bf CP}({\cal H})$.
Now assume that $Z^k\neq 0$, and define $z^j_{(k)}= Z^j/Z^k$ for $j\neq k$. Then 
$\sum_{j\neq k}^{\infty}\vert z^j_{(k)}\vert ^2 < \infty$ for every fixed 
value of $k$. As $j\neq k$ varies, these $z^j_{(k)}$ cover one copy of ${\cal H}$ 
that we denote by ${\cal U}_k$. The open set ${\cal U}_k$, endowed with 
the coordinate functions $z^j_{(k)}$, $j=1,2, \ldots \check k, \ldots$, 
where a check over an index indicates omission, provides a holomorphic coordinate chart 
on ${\bf CP}({\cal H})$ for every fixed $k$.
A holomorphic atlas is obtained as the collection of all pairs  $({\cal U}_k, 
z_{(k)})$, for $k=1, 2, \ldots$ There are nonempty $f$--fold overlaps $\cap_{m=1}^f {\cal U}_m$ 
for all values of $f=1, 2, \ldots$ Holomorphic tangent vectors and holomorphic 1--forms 
on ${\cal U}_k$ are spanned by $\partial/\partial z^j_{(k)}$ and ${\rm d}z^j_{(k)}$. 
When $f=2$ above, these vectors and covectors transform according to an 
(infinite--dimensional) jacobian matrix and its transpose.

${\bf CP}({\cal H})$ is a K\"ahler manifold. On the coordinate chart $({\cal 
U}_k, z_{(k)})$, the K\"ahler potential reads
\begin{equation}
K(z_{(k)}, {\bar z}_{(k)})=
\log{\left(1 + \sum_{j\neq k}^{\infty} z^j_{(k)} {\bar z}^j_{(k)}\right)},
\label{fubst}
\end{equation} 
and the corresponding metric ${\rm d}s^2_K$ reads on this chart
\begin{equation}
{\rm d}s^2_K = \sum_{m,n\neq k}^{\infty}{\partial^2 K(z_{(k)}, {\bar z}_{(k)})\over
\partial z_{(k)}^m \partial {\bar z}_{(k)}^n}\, {\rm d}z_{(k)}^m{\rm d}{\bar z}_{(k)}^n.
\label{rrdd}
\end{equation}
Being infinite--dimensional, ${\bf CP}({\cal H})$ is noncompact. It is simply connected:
\begin{equation}
\pi_1\left({\bf CP}({\cal H})\right)=0.
\label{cphx}
\end{equation}
Its Picard group is the group of integers:
\begin{equation}
{\rm Pic}\, ({\bf CP}({\cal H}))={\bf Z}.
\label{ppic}
\end{equation}
It has trivial homology in odd real dimension,
\begin{equation}
H_{2k+1}\left({\bf CP}({\cal H}), {\bf Z}\right)=0,\qquad k=0,1,\ldots,
\label{cotri}
\end{equation}
while it is nontrivial in even dimension,
\begin{equation}
H_{2k}\left({\bf CP}({\cal H}), {\bf Z}\right)={\bf Z}, \qquad k=0,1,\ldots
\label{pelotudo}
\end{equation}

\section{Quantisation of ${\bf CP}({\cal H})$}\label{cepehache}

By eqn. (\ref{ppic}), for each integer $l\in{\bf Z}$ there exists one equivalence class 
of holomorphic lines bundles over ${\bf CP}({\cal H})$; using the notations of ref. \cite{PQM} 
let us denote it by $N_l({\bf CP}({\cal H}))$. For $l\neq 0$ this bundle
is nontrivial; its fibre ${\bf C}$ is generated by the vacuum state $\vert 0\rangle_l$.
Let $A^{\dagger}_j(k)$, $A_j(k)$, $j\neq k$, 
be creation and annihilation operators on the chart ${\cal U}_k$, for $k$ fixed. 
We can now construct the ${\cal QH}_l$--bundle over 
${\bf CP}({\cal H})$. To this end we will describe the fibre over each coordinate chart 
${\cal U}_k$, plus the transition functions on the 2--fold overlaps ${\cal U}_k\cap {\cal U}_m$, 
for all $k\neq m$. 

The Hilbert--space fibre over ${\cal U}_k$ is ${\cal H}$ itself, the latter being the 
${\bf C}$--linear span of the infinite set of linearly independent vectors 
\begin{equation}
\vert 0(k)\rangle_l,\qquad A^{\dagger}_j(k)\vert 0(k)\rangle_l,\qquad 
j=1, 2, \ldots, \check k, \ldots
\label{nnumm}
\end{equation}
Reasoning as in ref. \cite{PQM} one proves 
that, on the 2--fold overlaps ${\cal U}_k\cap {\cal U}_m$, the fibre ${\cal H}$ can be chosen 
in either of two equivalent ways.  ${\cal H}$ is either the ${\bf C}$--linear span of the vectors 
$\vert 0(k)\rangle_l$, $A^{\dagger}_j(k)\vert 0(k)\rangle_l$, for
$j=1, 2, \ldots, \check k, \ldots$, or the ${\bf C}$--linear span of the vectors $\vert
0(m)\rangle_l$, 
$A^{\dagger}_j(m)\vert 0(m)\rangle_l$, for $j=1, 2, \ldots, \check 
m, \ldots$

Arguments identical to those of ref. \cite{PQM} prove that all the states of eqn. (\ref{nnumm}), 
except the vacuum $\vert 0(k)\rangle_l$, are (co)tangent vectors to ${\bf CP}({\cal H})$ on the 
chart ${\cal U}_k$, and thus transition functions are the sum of two
parts. One is a phase factor accounting for the transformation of
$\vert 0(k)\rangle_l$; the other one is a jacobian matrix. 
Identifying the tangent and cotangent bundles we can write
\begin{equation}
{\cal QH}_l({\bf CP}({\cal H})) = N_l({\bf CP}({\cal H})) \oplus T({\bf CP}({\cal H})).
\label{quhache}
\end{equation}

\section{Quantisation of ${\cal C}$}\label{imbbdd}

Sections \ref{bemet}, \ref{fedemb} present a summary, drawn from ref. \cite{CPINF}, 
on how to embed ${\cal C}$ holomorphically within ${\bf CP}({\cal H})$. 
This procedure is applied in section \ref{qqcccepeh} in order to quantise ${\cal C}$.
For background material see, {\it e.g.}, ref. \cite{PERELOMOV}.

\subsection{The Bergman metric on ${\cal C}$}\label{bemet}

Denote by ${\cal F}$ the set of holomorphic, square--integrable $n$--forms on ${\cal C}$.
${\cal F}$ is a separable, complex Hilbert space (finite--dimensional when ${\cal C}$ is compact).
Let $h_1, h_2, \ldots $ denote a complete orthonormal basis for ${\cal F}$, and let $z$ be (local) 
holomorphic coordinates on ${\cal C}$. Then
\begin{equation}
{\cal K}(z, \bar w)=\sum_{j=1}^{\infty}h_j(z)\wedge\bar h_j(\bar w)
\label{kaka}
\end{equation}
is a holomorphic $2n$--form on ${\cal C}\times\bar {\cal C}$, where $\bar {\cal C}$
is complex manifold conjugate to ${\cal C}$. The form ${\cal K}(z,\bar w)$ is 
independent of the choice of an orthonormal basis for ${\cal F}$; it is called 
the {\it kernel form}\/ of ${\cal C}$. If $\bar z$ is the point of $\bar {\cal C}$ 
corresponding to a point $z\in{\cal C}$, the set of pairs $(z,\bar 
z)\in {\cal C}\times \bar {\cal C}$ is naturally identified with ${\cal 
M}$. In this way ${\cal K}(z, \bar z)$ can be considered as a $2n$--form on 
${\cal C}$. One can prove that ${\cal K}(z, \bar z)$ is invariant under the group 
of holomorphic transformations of ${\cal C}$.

Next assume that, given any point $z\in{\cal C}$, there exists an 
$f\in{\cal F}$ such that $f(z)\neq 0$. That is, the kernel form 
${\cal K}(z,\bar z)$ of ${\cal C}$ is everywhere nonzero on ${\cal C}$:
\begin{equation}
{\cal K}(z,\bar z)\neq 0,\qquad \forall z\in{\cal C}.
\label{knz}
\end{equation}
Let us write, in local holomorphic coordinates $z^j$ on ${\cal C}$,  $j=1,\ldots, n$,
\begin{equation}
{\cal K}(z,\bar z)={\bf k}(z, \bar z)\,{\rm d}z^1\wedge\ldots\wedge{\rm d}z^n\wedge{\rm d}\bar 
z^1\wedge\ldots\wedge{\rm d}\bar z^n,
\label{kicco}
\end{equation}
for a certain everywhere nonzero function ${\bf k}(z, \bar z)$. 
Define a hermitean form ${\rm d}s^2_B$
\begin{equation}
{\rm d}s^2_B=\sum_{j,k=1}^n {\partial^2 \log {\bf k}\over \partial z^j \bar z^k}
\,{\rm d}z^j {\rm d}\bar z^k.
\label{bbmmff}
\end{equation}
One can prove that ${\rm d}s^2_B$ is independent of the choice of coordinates 
on ${\cal C}$. Moreover, it is positive semidefinite and invariant under 
the holomorphic transformations of ${\cal C}$.

Let us make the additional assumption that ${\cal C}$ is such that ${\rm d}s^2_B$ 
is positive definite,
\begin{equation}
{\rm d}s^2_B>0.
\label{posdef}
\end{equation}
Then ${\rm d}s^2_B$ defines a (K\"ahler) metric  called the {\it Bergman metric} 
on ${\cal C}$.

\subsection{Embedding ${\cal C}$ within ${\bf CP}({\cal H})$}\label{fedemb}

Let ${\cal H}$ be the Hilbert space dual to ${\cal F}$. Given $f\in {\cal F}$, let its expansion 
in local coordinates be
\begin{equation}
f={\bf f}\,{\rm d}z^1\wedge\ldots\wedge{\rm d}z^n,
\label{ppxx}
\end{equation}
for a certain function ${\bf f}$. Let $\iota'$ denote the mapping that sends $z\in {\cal C}$ 
into $\iota'(z)\in{\cal H}$ defined by
\begin{equation}
\langle\iota'(z)|f\rangle = {\bf f}(z).
\label{hhaa}
\end{equation}
Then $\iota'(z)\neq 0$ for all $z\in{\cal C}$ if and only if property (\ref{knz}) holds. 
Assuming that the latter is satisfied, and denoting by $p'$ the natural projection from 
${\cal H}-\{0\}$ onto ${\bf CP}({\cal H})$, the composite map $\iota = p'\circ\iota'$ 
\begin{equation}
\iota\colon{\cal C}\rightarrow{\bf CP}({\cal H})
\label{ccmmppp}
\end{equation}
is well defined on ${\cal C}$, independent of the coordinates, and holomorphic. 

One can prove the following results. When property (\ref{knz}) is true, the 
quadratic differential form ${\rm d}s^2_B$ of eqn. (\ref{bbmmff}) is the pullback, 
by $\iota$, of the canonical K\"ahler metric ${\rm d}s^2_K$ of eqn. (\ref{rrdd}):
\begin{equation}
{\rm d}s^2_B = \iota^*({\rm d}s^2_K).
\label{ppbbck}
\end{equation}
Moreover, the differential of $\iota$ is nonsingular at every point 
of ${\cal C}$ if and only if property (\ref{posdef}) is satisfied. 
These two results give us a geometric interpretation of the Bergman metric. 
Namely, if properties (\ref{knz}) and (\ref{posdef}) hold, 
then $\iota$ is an isometric immersion of ${\cal C}$ into ${\bf CP}({\cal H})$.

The map $\iota$ is locally one--to--one in the sense that every point of ${\cal C}$ 
has a neighbourhood that is mapped injectively into ${\bf CP}({\cal H})$. 
However, $\iota$ is not necessarily injective in the large. Conditions can 
be found that ensure injectivity of $\iota$ in the large. Assume 
that, if $z$, $z'$ are any two distinct points of ${\cal C}$, an 
$f\in{\cal F}$ can be found such that
\begin{equation}
f(z)\neq 0,\qquad f(z')=0.
\label{cdtn}
\end{equation}
Then $\iota$ is injective. Therefore, if ${\cal C}$ satisfies assumptions (\ref{knz}), 
(\ref{posdef}) and (\ref{cdtn}), it can be holomorphically and isometrically embedded 
into ${\bf CP}({\cal H})$.

\subsection{Quantisation of ${\cal C}$ as a submanifold of ${\bf CP}({\cal H})$}
\label{qqcccepeh}

Finally we quantise a noncompact ${\cal C}$ with infinite symplectic volume.
At any point of ${\cal C}$ there are only $n$ linearly independent, holomorphic tangent vectors. 
Applied to ${\cal C}$, the construction of ref. \cite{PQM} provides a finite--dimensional 
${\cal H}$, contrary to eqn. (\ref{isxvcc}). Therefore we try an alternative route.

We need an infinite--dimensional ${\cal QH}$--bundle over ${\cal C}$. 
For this purpose we assume embedding ${\cal C}$ holomorphically 
and injectively within ${\bf CP}({\cal H})$ as in eqn. (\ref{ccmmppp}). 
Then the bundle ${\cal QH}_l({\bf CP}({\cal H}))$ of eqn. (\ref{quhache}) can 
be pulled back to ${\cal C}$ by the embedding $\iota$. We take this {\it to define}\/
the bundle ${\cal QH}_l({\cal C})$:
\begin{equation}
{\cal QH}_l({\cal C}) = \iota^* {\cal QH}_l({\bf CP}({\cal H})).
\label{ppbbac}
\end{equation}
Even if ${\cal QH}_l({\bf CP}({\cal H}))$ were trivial (which it is not
for $l\neq 0$), it might contain nonflat (hence nontrivial) 
subbundles, thus allowing for nontrivial dualities.

A detailed analysis of ${\cal QH}_l({\cal C})$ requires specifying ${\cal C}$ explicitly.
However some properties can be stated in general. Thus, {\it e.g.},
the kernel form is the quantum--mechanical propagator. On ${\bf C}^n$ it reads 
\begin{equation}
{\cal K}_{{\bf C}^n}(z, \bar z) = N \,{\rm exp}\left({\rm i}\sum_{j=1}^n\bar z^j z^j\right)
{\rm d}z^1\wedge\ldots\wedge{\rm d}z^n\wedge{\rm d}\bar z^1\wedge\ldots\wedge{\rm d}\bar z^n,
\label{ceene}
\end{equation}
where $N$ is some normalisation. The Bergman metric (\ref{bbmmff}) derived from this kernel 
is the standard Hermitean metric on ${\bf C}^n$. The embedding $\iota$ naturally relates physical 
information (the propagator) and geometric information (the metric on ${\cal C}$). In retrospective, 
this justifies our quantisation of ${\cal C}$ by embedding it within ${\bf CP}({\cal H})$. 
An important physical input is the relation between the symplectic volume of ${\cal C}$ 
and the number of linearly independent quantum states.  It would be 
interesting to enquire into a possible geometric origin for this fact.

{\bf Acknowledgements}

It is a great pleasure to thank J. de Azc\'arraga, U. Bruzzo and M. Schlichenmaier 
for encouragement and discussions. This work has been partially supported by research grant 
BFM2002--03681 from Ministerio de Ciencia y Tecnolog\'{\i}a and EU FEDER funds.

\end{document}